\documentclass[journal]{IEEEtran}

\usepackage{cite}
\usepackage{amsmath,amssymb,amsfonts}
\usepackage{algorithm,algorithmic}
\usepackage{graphicx}
\usepackage{textcomp}
\usepackage{color,xcolor}
\usepackage{bm}
\usepackage{subfigure}
\usepackage{multirow}
\usepackage[normalem]{ulem}
\usepackage{tabularx}
\usepackage{xpatch}
\usepackage{soul,ulem}
\usepackage{verbatim}
\usepackage{geometry}

\ifCLASSINFOpdf
\else
\fi

\begin{document}

%

\title{Unified Far-Field and Near-Field in Holographic MIMO: A Wavenumber-Domain Perspective}

\author{Yuanbin~Chen,~Xufeng~Guo,~Gui~Zhou,~\IEEEmembership{Member,~IEEE,}~Shi~Jin,~\IEEEmembership{Fellow,~IEEE,}\\Derrick~Wing~Kwan~Ng,~\IEEEmembership{Fellow,~IEEE,}~and~Zhaocheng~Wang,~\IEEEmembership{Fellow,~IEEE}
\vspace{-1em}

\thanks{
This work was supported by the National Natural Science
Foundation of China under Grant U22B2057. \textit{(Corresponding author: Zhaocheng Wang.)}
}

%
%
%
%




}

\maketitle

\begin{abstract}
This article conceives a unified representation for near-field and far-field holographic multiple-input multiple-output (HMIMO) channels, addressing a practical design dilemma: ``Why does the angular-domain representation no longer function effectively?" To answer this question, we pivot from the angular domain to the wavenumber domain and present a succinct overview of its underlying philosophy. In re-examining the Fourier plane-wave series expansion that recasts spherical propagation waves into a series of plane waves represented by Fourier harmonics, we characterize the HMIMO channel employing these Fourier harmonics having different wavenumbers. This approach, referred to as the wavenumebr-domain representation, facilitates a unified view across the far-field and the near-field. Furthermore, the limitations of the DFT basis are demonstrated when identifying the sparsity inherent to the HMIMO channel, motivating the development of a wavenumber-domain basis as an alternative. We then present some preliminary applications of the proposed wavenumber-domain basis in signal processing across both the far-field and near-field, along with several prospects for future HMIMO system designs based on the wavenumber domain.

\end{abstract}


%
\IEEEpeerreviewmaketitle

\section{Introduction}

The fifth-generation (5G) base stations (BSs), equipped with massive multiple-input multiple-output (mMIMO) capabilities, have marked their successful commercial footprints across various nations and regions. Amidst the continual evolution of technology, towards the emerging sixth-generation (6G) communications, the wireless communications community remains unwavering in its commitment to exploring and sculpting future MIMO technologies with qualities hitherto unimagined – extremely high spectral and energy efficiency, ultra-low latency, and seamless connectivity. In this endeavor, it is anticipated to encompass the possibility of not only utilizing larger array apertures but also arrays with densely spaced antenna elements. This approach will be implemented over the millimeter-wave (mmWave) spectrum and even terahertz frequencies, offering more flexibility in manipulating the wireless environment \cite{GE2}.

Indeed, under classical operating conditions (i.e., the assumption of plane-wave propagation), the distance of the communication link significantly exceeds the size of antenna aperture. However, this scenario changes intriguingly when the antenna size becomes comparable to the distance of the communication link, in which the operating conditions shift into the Fresnel region characterized by the near-field propagation. Under these circumstances, it becomes imperative to consider a spherical-wave model instead. Therefore, operations within the Fresnel region necessitate the examination of novel models that accommodate this regime, which, in turn, paves the way for uncharted possibilities to enhance the performance of communication systems.

\subsection{Holographic MIMO Towards 6G}

Going beyond the current mMIMO paradigm,  which solely relies on Shannon's theory, electromagnetic (EM) information theory, an elegant blending of Maxwell and Shannon's principles, is poised to shine brilliantly in 6G wireless systems due to its capability to characterize the physical fundamental of EM wave propagation. In this sense, a holistic control over the EM fields sensed/generated by antennas should be realized. This conceptualizes the essence of the holographic MIMO (HMIMO) \cite{Holo-105}.
Viewed through HMIMO's physical fundamentals, the advancement in meta-materials provides a viable solution for creating highly flexible antenna elements, allowing the manipulation of the EM field with precise control over amplitude and phase at an unprecedented level. Meanwhile, meta-surfaces, abstracted form specific meta-materials implementations, can be modeled as continuous arrays composed to infinitesimally small antenna elements, essentially forming a nearly continuous aperture. 
With the theoretical integration of numerous antenna elements in a continuous aperture, one can treat this as the asymptotic limit of mMIMO. However, the increasing size of these surfaces poses a challenge, i.e., wireless propagation can readily occur within an expanded near-field region when utilizing an array operating at millimeter-wave or terahertz frequencies. Thus, the plane-wave assumptions may cease to retain their validity.

\subsection{State-of-the-Art in HMIMO Channel Representations}
Upon examining the existing literature on HMIMO, we identify a twofold lack of uniformity regarding HMIMO channel representations: i) the absence of a unified approach in HMIMO channel modeling and ii) different  representations of near-field and far-field regimes. Firstly, as regards the HMIMO channel modeling, the state-of-the-art research tends to bifurcate into two different kinds. The first kind revolves around stochastic EM channels based on plane-wave expansion \cite{Holo-4,Holo-2, guo-ICC24,Holo-2-OJCOM}, aiming to restructure the channel into a spatially stationary one that is instrumental in capturing the essence of EM propagation by describing the angular characteristics between source and receiver. This is achieved through the application of Fourier series analyses. The second kind of structure is based on the parameterized physical channel, characterized by the classical multi-path representation \cite{Holo-15,chen-jsac3,guo-tvt}. This channel representation, derived from the clustered delay line (CDL) model, aims to encapsulate significant paths through angle-distance parameters. Despite these differences, we deem that this pair of models fundamentally converge in their focus on the impact of clusters in the scattering environment on channel sparsity.

Secondly, the non-uniformity in near-field and far-field representations stands as another serious concern. Indeed, different representations of the far-field and near-field channels come from different approximations of the channel response. Specifically, these include the Fraunhofer approximation in the far-field\cite{GE2}, the Fresnel approximation in the near-field \cite{guo-ICC24,chen-jsac3}, and the Fourier plane-wave series approximation \cite{Holo-2,guo-ICC24}. 
The Fourier plane-wave series approximation is determined by the plane waves characterized by different wavenumbers, exhibiting beneficial distance-independent properties.  However, the discussion in \cite{Holo-2,guo-ICC24} particularly restricts its application to far-field scenarios based on certain assumptions, such as the ignored evanescent waves. 
Additionally, there have been some fledgling attempts at integrating the representations of the far-field and near-field, which is in essence a combination of the Fraunhofer and Fresnel approximation techniques \cite{sunshu}. Despite claims of applicability in both far-field and near-field, the modeling and techniques employed still require grappling with an additional distance parameter. Therefore, inspired by the distance-independent feature of the Fourier plane-wave series approximation, further investigations are required to achieve a unified channel representation for both the far-field and near-field of HMIMO systems.

\subsection{Challenges and Motivations}

As mentioned in the state-of-the-art, this pair of non-uniformities poses significant challenges for the accurate characterization of near-field and far-field HMIMO channels. It is observed that the HMIMO channel modeling based on Fourier plane-wave series expansion is gaining wider recognition, since such a kind of structure has the potential to describe the channel response between a source and a receiver under arbitrary scattering conditions \cite{Holo-2-OJCOM,Holo-3}. Furthermore, it is indisputable that although near-field analysis is indispensable, far-field communication still dominates a vast majority of application scenarios, rendering the near-field analysis more of an enhancement than a necessity. Therefore, the aim of this article is to develop a unified framework for effectively describing the HMIMO channel in both near-field and far-field contexts, while unveiling the potential sparsity thereof. In this respect, the following two key questions deserve our additional attention.

\textbf{Question 1:} \textit{Why does the angular-domain representation work inefficiently in revealing the sparsity of near-field and far-field HMIMO channels?}
In traditional mMIMO channels, the use of a discrete Fourier Transform (DFT) as a sparsifying basis facilitates the transformation of far-field channels into their angular-domain counterparts, based upon which significant paths corresponding to clusters can be identified. This is instrumental in recovering channel parameters at reduced complexity and signaling overhead. However, the spherical-wave assumption in the near field introduces spatial non-stationarity, which is in contradiction with a spatially stationary system described by the Fourier plane-wave series expansion in \cite{Holo-2}. Additionally, even in the far-field, the dense antenna spacing in HMIMO connotes that the DFT basis is unable to capture complete significant power, which naturally leads to the second question.

\textbf{Question 2:} \textit{What is the fundamental impact of densely spaced antenna elements, e.g., with quarter-wavelength spacing, on the sparse representation of HMIMO channels?} In the case of quater-wavelength antenna spacing, the DFT basis fails to fully capture the spatial information encompassed by the HMIMO channel that is continuously represented by the Fourier series. This limitation often leads to the misidentification of non-significant power as significant, resulting in the angular power spectrum leakage across different angles. This phenomenon is referred to as power leakage. Employing the unified framework to be introduced in this work, a wavenumber-domain representation, when exploiting a sampling interval corresponding to a quarter of the wavelength, demonstrates a marked ability to accurately identify significant angular power. The conceived unified framework aims at effectively addressing ambiguous detection in HMIMO channel analysis.

Geared towards the pair of questions outlined above, we conceive a unified framework for characterizing the HMIMO channel in both near-field and far-field contexts. Given the fact that EM waves always propagate as spherical waves in free space, a spherical wave can be expanded into a linear superposition of a series of plane waves, parameterized by wavenumbers, thus constructing a spatial stationary system.  This kind of expansion is applicable regardless of the distance between the source and receiver, thereby laying a concrete foundation for the development of a unified representation of HMIMO channels in both near-field and far-field contexts. Therefore, these plane-wave functions take the form of Fourier series, in which a selective number of functions serve as an orthonormal series expansion, constituting the Fourier harmonic (FH) basis in the wavenumber domain. This is also referred to as the waveumber-domain basis. Furthermore, the wavenumber-domain basis facilitates the exposition of sparsity in HMIMO channels. We demonstrate that HMIMO channels, when filtered by the wavenumber-domain basis, exhibit clustered sparsity in terms of the angular power spectrum in both near-field and far-field scenarios. 
Given the beneficial attributes of the wavenumber-domain basis, we explore its applications in signal processing for HMIMO systems, including sparse channel estimations and codebook design. Finally, several prospective avenues are showcased for future HMIMO system designs grounded in the wavenumber domain.

\section{What is the Wavenumber Domain?}

Although the concept of wavenumber is not new, its application in the field of signal processing in MIMO systems is non-intuitive and challenging to grasp. Hence, this section aims to present some helpful insights into the fundamentals the wavenumber domain.

Prior to exploring the secrets of the wavenumber domain, it is instructive to reacquaint ourselves with the classical Fourier transform -- a foundational tool that facilitates a deeper understanding of the fundamentals of the wavenumber domain.
The one-dimensional Fourier transform employs the continuous function $f \left( t\right) $  as the pre-image function, with Fourier harmonics $ {e^{ - j\omega t}}$ serving as the kernel function, and the spectrum function $F\left( \omega  \right)$ acting  as the image function after an integral operation, where $t$ represents time and $\omega$ denotes the angular frequency. From a dimensional perspective, a signal in the time domain is measured in seconds, whereas its image function is measured in the angular frequency $\omega$, with dimensions of $2\pi  \cdot {{\text{s}}^{ - 1}}$. Similarly, when the spatial-domain signal measured in meters (m) serves as the pre-image function, the dimension of its image function in the frequency domain should be $2\pi \cdot \text{m}^{-1}$. Note that in the realm of electromagnetism, the wavenumber $k = 2\pi / \lambda$ is the only physical quantity that corresponds to this dimension. This observation indicates that the wavenumber may serve as a link between the spatial-domain signal and the frequency-domain signal. The spatial-domain signal after the DFT is in essence the superposition of a series of Fourier harmonics (FHs) characterized by different wavenumbers instead of angles.

\begin{figure}[t]
	\centering
	\includegraphics[width=0.5\textwidth]{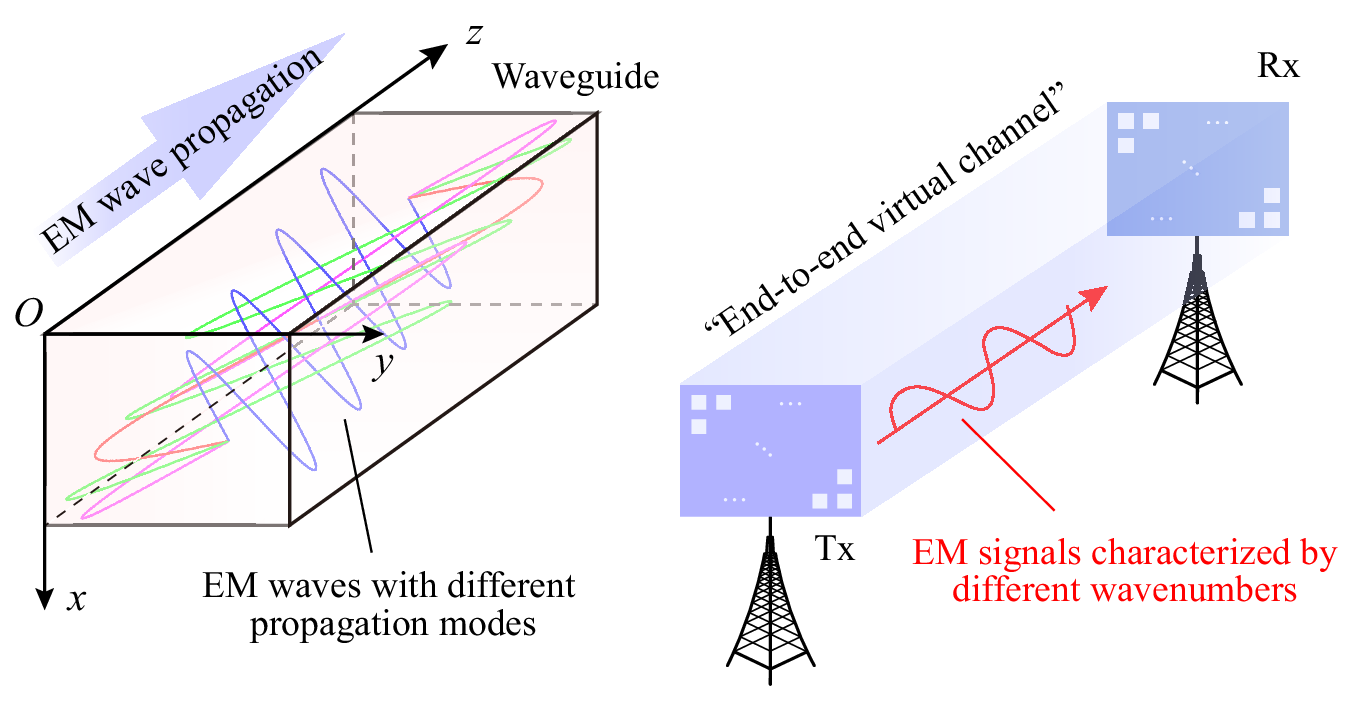}
	\caption{Illustration of the EM signals characterized by different wavenumbers.} \label{wavenumber}
			\vspace{-0.5em}
\end{figure}

For a more intuitive grasp of the wavenumber domain, we borrow some terminologies from microwave theory and draw similarities to signal processing in MIMO systems. As illustrated in Fig.~\ref{wavenumber}, considering a rectangular metal waveguide, if the cross-section is taken as the $xOy$ plane and the direction of the waveguide is the positive $z$-axis, then the wavenumbers in the $x$, $y$, and $z$ directions are denoted by $k_x$, $k_y$, and $k_z$, respectively. The wave propagating in the waveguide is composed of a superposition of EM waves determined by a finite number of propagation modes. Similarly, in a MIMO system, if the propagation channel spanning from the transmitter to the receiver can be ragarded as a ``waveguide", the end-to-end signal can be treated as a superposition of EM waves characterized by different propagation modes (wavenumbers) propagating in the ``rectangular waveguide" (propagation channel). Therefore, the essence of the wavenumber-domain representation is employing a series of FHs characterized by different wavenumbers to represent the EM signals. The wavenumber domain can also be understood as the spatial-frequency domain \cite{book-EM}, exploiting the wavenumber to characterize EM signals.

\section{HMIMO Channels in the Wavenumber Domain}




The wireless channel is an abstract concept, and let us consider a general situation in order to better understand its fundamentals. Given a wireless environment consisting of a source, $\mathbf{s}$, and a receiver, $\mathbf{r}$, the electric field observed at the receiver is inherently associated with the electric-current volume density at the source. This relationship encompasses the function $h \left(  \mathbf{s},\mathbf{r} \right) $, representing the spatial impulse response of the channel at $\mathbf{r}$ in response to an impulsive electric current introduced at $\mathbf{s}$. 

For a HMIMO case where $\mathbf{s}$ and $\mathbf{r}$ are array-based, the complex signals $h \left(  \mathbf{s},\mathbf{r} \right) $ stimulated by the source array serve as the entries of the channel matrix.
More explicitly, based on the Helmholtz equation derived from Maxwell's equations, the channel response sensed by the antenna element positioned at $\mathbf{r} = \left( {r_x, r_y, r_z} \right) $, is in essence the Green's function field radiated by a source antenna element positioned at $\mathbf{s} = \left( {s_x, s_y, s_z} \right) $. Given the intricacies of the Green's function, communication  theorists have consistently pursued appropriate approximations that can be easily analyzed. 
To this end, this section commences by briefly revisiting three techniques for approximating the channel response.
Then, based on the Fourier plane-wave series approximation,
we show the wavenumber-domain representation for HMIMO channels.

\begin{figure*}[t]
	\centering
	\includegraphics[width=0.9\textwidth]{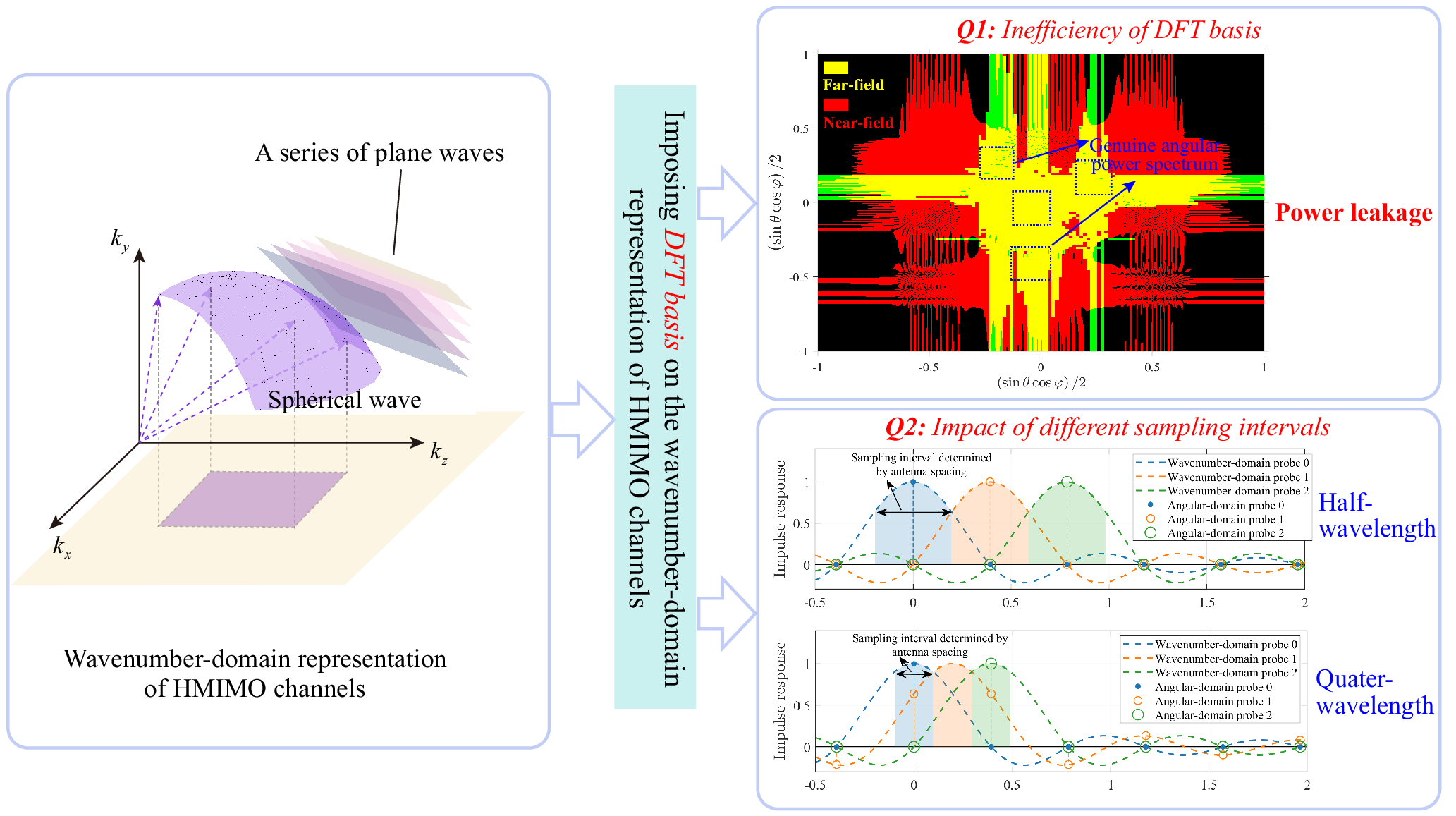}
	\caption{An illustration of the wavenumber-domain representation and the challenges when employing the DFT basis.} \label{challenge}
			\vspace{-0.5em}
\end{figure*}

\subsection{Revisiting Approximation Techniques for Channel Response}

\subsubsection{Fraunhofer Approximation}
The physical significance of the Fraunhofer approximation lies in that the spherical wavefronts of the EM wave are treated as the planar counterparts when the distance between the source $\bf{r}$ and the receiver $\bf{r}$ exceeds Rayleigh distance. However, the Fraunhofer approximation exhibits inaccuracies in the near-field region, particularly for XL-MIMO and HMIMO systems.


\subsubsection{Fresnel Approximation}
The accuracy of the Fresnel approximation outperforms that of the Fraunhofer approximation in the near-field scenario, owing to a pair of reasons. Firstly, from a geometric perspective, the Fraunhofer approximation treats the spherical wavefronts as planar ones, whereas the Fresnel approximation models them as parabolic wavefronts.
Secondly, in an algebraic view, the Fraunhofer approximation limits its Taylor series expansion to the first order, while the Fresnel approach extends to the second order, resulting in fewer residual terms. These two-fold attributes endow the Fresnel approximation a higher degree of accuracy. Despite its great accuracy, the channel representation based on the Fresnel approximation entails high computational complexity due to the presence of nonlinear terms. The accuracy might be offset by the errors induced by the intricate numerical calculations.

\subsubsection{Fourier Plane-Wave Series Approximation}
It is evident that the two approximation methods mentioned above are the approximations of the Green's function, aiming for achieving more tractable mathematical expressions. However, the resultant errors are determined by the distance between $\bf{r}$ and $\bf{s}$. To obtain a unified channel response that is applicable at any distance, a viable approach is to employ the Weyl Identity to transform the channel response into an integral over the wavenumbers \cite{Holo-2-OJCOM}. Specifically, by harnessing the Weyl identity, the channel response $h \left(  \mathbf{s},\mathbf{r} \right) $ takes the form of a selective number of plane waves that are radiated by the source $\mathbf{s}$, irrespective of the distance between the source $\mathbf{s}$ and the receiver $\mathbf{r}$. Furthermore, \cite[Eq.~(21)]{Holo-2} provides an exact analytic expression of the Fourier plane-wave series approximation. In this formula, we learn that the physical meaning of the FH coefficient represents the complex gain along the propagation direction, and integrating the wavenumber over the specific integral area constitutes the wavenumber domain i.e., the lattice ellipses specified in \cite[Eqs.~(17)\&(18)]{Holo-2}.


\subsection{Wavenumber-Domain Representation for HMIMO}

By leveraging the Fourier plane-wave series approximation, the EM waves are uniformly represented as a linear superposition of a series of linear and stationary plane waves {\color{blue}with respect to the wavelength $\lambda$}. Based on this, the channel response can be precisely parameterized by wavenumbers, which is referred to as the \textit{wavenumber-domain representation}.
When the HMIMO channel is represented in this manner, it encapsulates the source/receiver response that maps the excitation current at any point to the corresponding propagation direction of the EM radiated field from the source/receiver viewpoint. It also includes the angle-wise response that associates with the direction of the source to that of the receiver. This implies that regardless of the distance between the source and receiver, our focus is on the angle-related plane-wave directions (characterized by angles) and the power associated with these directions, namely the angular power spectrum.

Given that the antenna spacing being less than half a wavelength results in mutual couplings among antenna elements. To address this issue, a
polarized EM channel model for HMIMO is proposed in \cite{Holo_22-JSAC}. These adjustments are achieved by left-multiplying and right-multiplying the FH coefficient matrix with matrices that encapsulate the polarization leakage of the propagation path, antenna efficiency, and the embedded element directivity patterns. The embedded element directivity patterns specifically characterize the antenna pattern distortion induced by mutual couplings. It
thus provides a more exact representation in the presence of physical constraints, as detailed in \cite[Eq.~(33)]{Holo_22-JSAC}.
However, these refinements neither alter the essence of the wavenumber-domain representation, nor will they affect the sparsity structure of the HMIMO channels. This is because these physical constraints -- determined by matrices -- are established once the antenna array is manufactured. Given these physical constraints, achieving sparsity in the refined channel only requires performing some appropriate mathematical tricks, such as left-multiplying and right-multiplying by the pseudo-inverse of these matrices.

\section{Wavenumber-Domain Basis}

In this section, we conceive a novel wavenumber-domain basis in order to uniformly represent near-field and far-field HMIMO channels. Before proceeding, let us first examine why the DFT basis fails to accurately identify the sparsity of HMIMO channels.

\subsection{Limitations of DFT Basis}

Recall that the success of a DFT basis in accurately identifying significant angles of the far-field channel can be attributed to its stringent orthogonality and the spatial stationarity based on a plane-wave model.
From the perspective of EM wave propagation, these significant angles specifically correspond to those FHs that have the strongest EM impulse responses. This implies that there may exist a mapping relationship between the wavenumber and the angle. Having said that, given that the definition domain of the wavenumber is different from the angular counterpart, some wavenumbers determined by the DFT atoms do not have their corresponding angles.
As a result, one may observe a power leakage in the spectrum after imposing the DFT basis to a channel represented in the wavenumber domain, as portrayed in Fig.~\ref{challenge}. 



\subsubsection{Far-Field Case}
In Fig.~\ref{challenge}, the HMIMO channel is generated according to the channel model presented in \cite{Holo-2,guo-ICC24}, and 
herein we consider four clusters in the propagation environment. An observation of dotted blue squares at the center may represent the genuine angular power spectrum for the clusters. However, the angular power spectrum seems to disperse outward in a manner reminiscent of ``leakage" or ``spreading", which prevents us from observing the precise clustered sparsity inherent in the HMIMO channel. This is referred to as power leakage.

The power leakage can be attributed to the sampling mismatch presented in Fig.~\ref{challenge}. In particular, two distinct sampling techniques are employed: (i) the angular-domain probe characterized by a set of discrete points, and (ii) the wavenumber-domain probe characterized by a continuous Dirichlet kernel representation. Herein, we exploit three probes in the angular domain and wavenumber domain to scrutinize their individual sampling capabilities. We observe that in the case of half-wavelength antenna spacing, both non-zero and zero impulse responses align with the Dirichlet kernel's peaks and nulls. This alignment indicates that significant angular power spectrum can be accurately identified in the case of half-wavelength antenna spacing, regardless of the type of probes employed. By contrast, in the case of quarter-wavelength antenna spacing, the angular-domain probes fail to detect the true impulse response of angular power.  The refined sampling interval may result in mismatch between the genuine impulse response and the sampled value. More precisely, regarding the case of the quater-wavelength antenna spacing, the impulse response sampled by the angular-domain probe 1, as depicted in Fig.~\ref{challenge}, bears no relation to a significant angle. 
Even if the sampled impulse response is weak, it might be misinterpreted as a significant one when employing the angular-domain probe. This seemingly ``diligent” sampling is in fact a fruitless detection devoid of any physical significance. Therefore, such ambiguous detection is the fundamental cause of the power leakage.

\subsubsection{Near-Field Case}

\begin{figure}[t]
	\centering
	\includegraphics[width=0.4\textwidth]{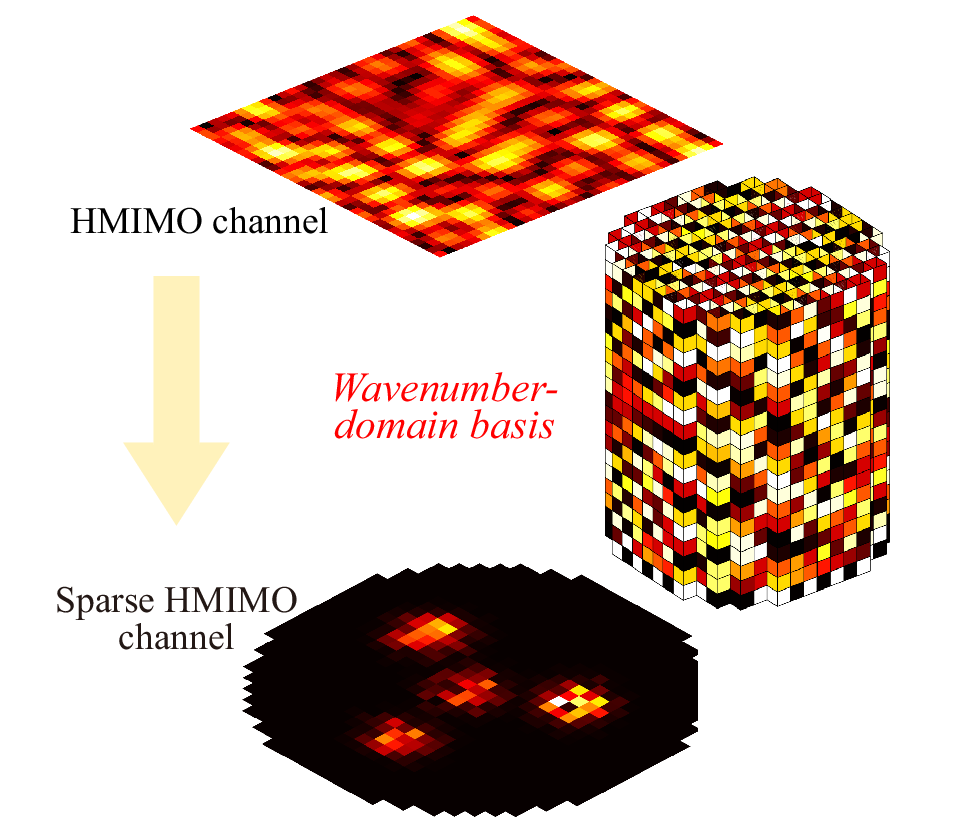}
	\caption{The implementation of the wavenumber-domain basis exposes the sparsity inherent in the HMIMO channel.} \label{WD_Basis}
\end{figure}

To examine the spatial non-stationary of HMIMO channels in the near-field, we continue to focus our attention on the channel model presented in \cite{guo-ICC24}, where the distance between the source and receiver is within the Rayleigh distance. When imposing the DFT basis on the near-field HMIMO channel, the angular power spectrum continues to extend beyond its central region, resulting in what we refer to as near-field power leakage. Furthermore, from Fig.~\ref{challenge}, it is difficult to identify the significant angular power spectrum associated with the limited number of scattering clusters present in the propagation environment. This observation can be ascribed not only to the sampling mismatches presented in Fig.~\ref{challenge}, but also to the spatial non-stationarity, as different  directions observed reveal that each point on the antenna array has its individual EM field. Thus, in the event of a spatial non-stationary state, the application of a DFT basis with uniform sampling capabilities further exacerbates the problem, resulting in more pronounced power leakage compared to far-field scenarios.

\subsection{FH-Based Wavenumber-Domain Basis}
The design of the wavenumber-domain basis is straightforward. Specifically, recall that the HMIMO channel response $h \left( {\mathbf{s}, \mathbf{r}}\right) $
is composed of a selective number of plane waves radiated by the source $\mathbf{s}$, regardless of the distance between $\mathbf{s}$ and $\mathbf{r}$. This operation engenders a sequence of orthogonal functions.
Selecting any function from this sequence allows it to serve as an orthonormal series expansion, also referred to as the FH-based wavenumber-domain basis. One may refer to our previous work in~\cite[Eq.~(9)]{guo-ICC24} for its analytical form. 
Note that following the assumptions made in~\cite{Holo-2}, the basis in \cite{guo-ICC24} excludes all waves (i.e., evanescent waves) that decay exponentially along the $z$-axis. This assumption is particularly applicable to the far-field case with small elevation angles. By contrast, in the near-field scenario, the evanescent waves should not be ignored \cite{Holo_27}. Therefore, we refine this approach such that the wavenumber-domain basis proposed in this article can accurately identify the sparsity of HMIMO channels in both near-field and far-field scenarios.

Fig.~\ref{WD_Basis} demonstrates HMIMO channel's angular power spectrum and its counterpart after wavenumber-domain basis filtering.
We employ a Von Mises-Fisher (VMF) distribution to model the angular power spectrum, in conjunction with parameters presented in 3GPP TR 38.901 \cite{3GPP-38901}. This approach effectively models a non-isotropic scattering environment, allowing for the angular power spectrum with a realistic and physically interpretable structure. 
Evidently, the sparsity of HMIMO channels can be consistently identified by employing the wavenumber-domain basis, irrespective of whether the observation is in the far-field or near-field. The sparsity observed from both the far-field and the near-field is identical, as transpired in Fig.~\ref{WD_Basis}. The efficiency of the considered wavenumber-domain basis lies in the ability to accurately characterize angles with wavenumbers having physical meanings, thus eliminating the power leakage. Furthermore, it reveals beneficial attributes of being independent from the sampling intervals determined by the antenna spacing, despite the antenna elements being more densely packed.

\section{Applications of Wavenumber-Domain Basis}

Given that users in actual systems are unaware of whether they are in the far-field or near-field, the wavenumber-domain paradigm serves as the cornerstone for signal processing at any transmitter-receiver distance.
In what follows, this section examines two key applications of the wavenumber-domain basis: sparse channel estimation and precoding. It demonstrates how the wavenumber-domain paradigm effectively unifies the far-field and near-field scenarios.

\subsection{Sparse Channel Estimation}

HMIMO channels tend to be sparse in the presence of the wavenumber-domain basis, facilitating the acquisition of their channel parameters. Although some classical compressive sensing techniques, e.g., orthogonal matching pursuit (OMP)~\cite{guo-tvt}, are capable of retrieving non-zero entries in the sparse matrix, they fall short in encapsulating the clustered sparsity in HMIMO channels.  Generally, the sparsity level of HMIMO channels exceeds the number of clusters. This results in the number of running loops in OMP being significantly greater than the number of clusters. 
Inspired by the elliptic shape of the wavenumber domain, the Markov Random Field (MRF)-based approach is capable of capturing the spatial correlation, e.g., a priori and a posteriori \cite{guo-ICC24}.



\begin{figure}[t]
	\centering
	\includegraphics[width=0.47\textwidth]{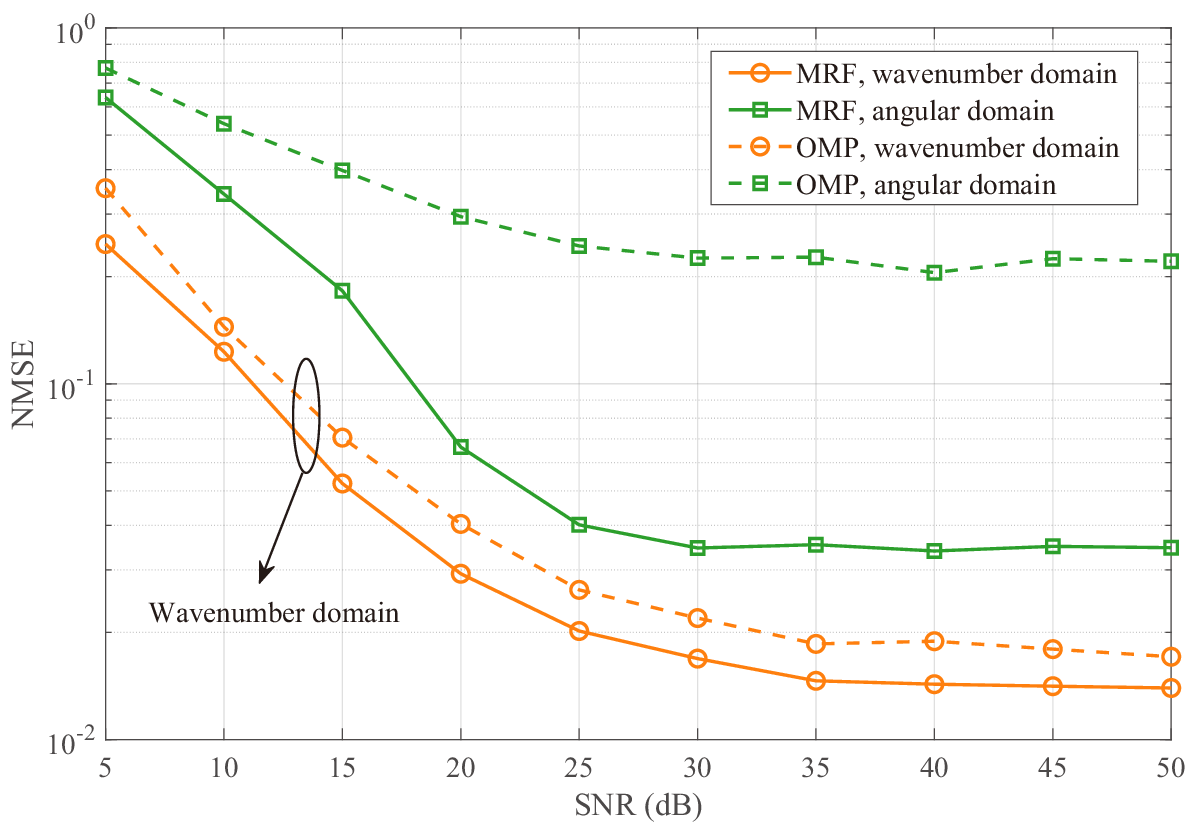}
	\caption{NMSE performance achieved by angular-domain and wavenumber-domain representation versus varying SNR.} \label{NMSE}
			\vspace{-1em}
\end{figure}

Fig.~\ref{NMSE} demonstrates the normalized mean square error (NMSE) performance versus the signal-to-noise (SNR). The classical algorithm, i.e., OMP, is employed as a benchmark, in order to underscore MRF's remarkable capability  in capturing the wavenumber-domain clustered sparsity. As it transpires, the considered MRF method outperforms the traditional OMP algorithm in terms of NMSE performance in both wavenumber and angular domains. 
Even in the low SNR regime, the MRF method still delivers favorable performance. Additionally, the saturation of NMSE performance achieved by all schemes is attributed to the intrinsic properties of the algorithm. Both algorithms employ an angular power threshold as a criterion to determine the presence of non-zero entries, in which the threshold is a paramount determinant of the estimation accuracy that the algorithm can achieve best~\cite{guo-tvt}.

\subsection{Codebook Design}
The wavenumber-domain basis can also serve as a codebook in HMIMO systems. For a preliminary attempt, we consider a single-user case to verify the effectiveness of the wavenumber-domain basis. We employ the refined HMIMO channel model presented in \cite{Holo_22-JSAC}. The base station is equipped with a uniform planar array having $129 \times 65$ antennas with an antenna spacing of $\lambda/4$, and the considered HMIMO system operates at $30$~GHz. This configuration results in a Rayleigh distance of $26.85$ meters. Furthermore, the DFT-based codebook is considered as a benchmark scheme.

\begin{figure}[t]
	\centering
	\includegraphics[width=0.46\textwidth]{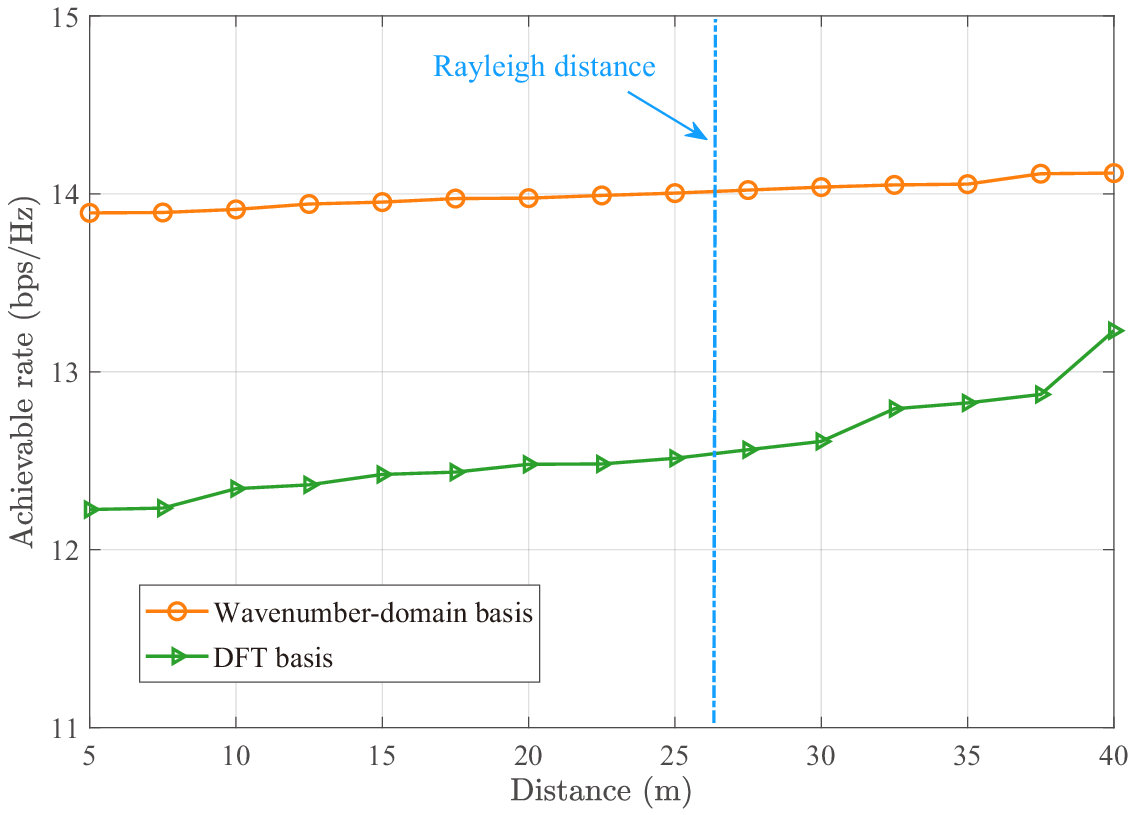}
	\caption{Achievable rate versus distance achieved by the wavenumber-domain basis and DFT basis.} \label{rate}
\end{figure}

Fig.~\ref{rate} shows the achievable rate versus the distance between the user and the base station. As it transpires, the achievable rate attained by the wavenumber-domain codebook outperforms that of DFT. In examination of reasons, recall that some wavenumbers determined by the DFT atoms do not have their corresponding angles, i.e., these wavenumbers are the high wavenumbers. Indeed, channel estimate errors result in the utilization of invalid codewords in DFT atoms for beamforming. The beams associated with these codewords attenuate exponentially, leading to performance erosion. Additionally, the wavenumber-domain scheme delivers a consistent achievable rate regardless of varying distance, while the achievable rate obtained by DFT exhibits fluctuations. This confirms the effectiveness of our proposed wavenumber-domain basis as a codebook, functioning well across both far-field and near-field scenarios.

\section{Future Directions and Concluding Remarks}

This article reveals the potential of HMIMO being unified in both near-field and far-field scenarios from the channel representation level. Although a customized framework based on the wavenumber-domain is conceived, there are still non-trivial challenges spanning from academic to the industry, from theoretical analyses to practical implementations, and from HMIMO-only technology to an intrinsically holography-capable wireless environment. Below, we provide several promising topics for future exploration.

\textbf{HMIMO Theoretical Fundamental in the Wavenumber Domain:} Shannon's classical information theory lays a solid foundation for contemporary wireless communications, adopting a mathematical approach that observes the wireless channel through the lens of a specific probability distribution.
However, its application encounters significant limitations when addressing the fundamental constraints of the HMIMO regime. This is primarily due to its inadequate consideration of the physical properties of EM wave propagation. This gap may hamper a direct application of Shannon’s theory in the context of HMIMO system.
Fortunately, the emerging concept of EM information theory, which integrates principles from both Shannon's and Maxwell's theories, is envisioned as an intriguing advancement in shaping the analysis and design for future wireless systems. 
Therefore, by blending Shannon's and Maxwell's classic theories under the proposed unified view in the wavenumber domain, an in-depth exploration of HMIMO may be achieved in order to go beyond the limits of current wireless communications.


\textbf{Multi-User HMIMO Precoding Using the Wavenumber-Domain Basis:} The effectiveness of the wavenumber-domain basis have been validated in the single-user case. However, the main bottleneck in enhancing the performance of multi-user HMIMO systems lies in the interference among users, while taking into account the antenna efficiency, antenna pattern, and the polarization leakage of all propagation paths. Therefore, the feasibility of using the wavenumber-domain basis for a multi-user codebook in HMIMO systems requires further investigations. Additionally, it is also an open issue whether an alternative architecture may outperform the existing hybrid precoding framework in efficiently handling the intricacies of HMIMO system signal processing.

\textbf{HMIMO Hardware Implementation Accommodating Wavenumber-Domain Representation:} Albeit significant advancements in meta-materials technologies, such as multi-beam antennas and waveguide-fed metasurface-based dynamic metasurface antennas (DMAs), there remains a substantial gap that has to be addressed to materialize truly versatile HMIMO arrays. This includes various unresolved issues regarding its practical implementations, e.g., waveguide attenuation with varying degrees as well as mutual coupling among antenna elements \cite{Holo-105}. Although it is possible to theoretically evaluate these imperfect factors leveraging mathematical techniques	as previously mentioned in this article, a fully digital beamforming architecture might be required  for achieving the expected performance, which is currently not feasible for HMIMO systems. Thus, further investigation and development of sophisticated hardware architectures for accommodating the wavenumber-domain representation are required for effectively addressing these challenges.


The evolution of holographic MIMO technologies introduces ever-increasing complexities but brings a host of possibilities for achieving an intelligent and endogenously holography-capable wireless propagation environment. This article addresses the challenges associated with applying traditional angular-domain representations to HMIMO channels and, in response, puts forward a unified framework that uniformly represents HMIMO channels across both near-field and far-field scenarios. The introduction of the FH basis, employing a quarter-wavelength sampling interval, facilitates the exposition of clustered sparsity in HMIMO channels in the absence of power leakage. It sharpens the identification of the angular power spectrum associated with each cluster in the scattering environment. Additionally, a case study is provided for sparse estimation, aimed at precisely recovering the angular power spectrum by fully capturing the clustered sparsity of HMIMO channels. 


\ifCLASSOPTIONcaptionsoff
  \newpage
\fi

\bibliographystyle{IEEEtran}
\bibliography{ref_Uni_Holo}

\begin{thebibliography}{10}
\providecommand{\url}[1]{#1}
\csname url@samestyle\endcsname
\providecommand{\newblock}{\relax}
\providecommand{\bibinfo}[2]{#2}
\providecommand{\BIBentrySTDinterwordspacing}{\spaceskip=0pt\relax}
\providecommand{\BIBentryALTinterwordstretchfactor}{4}
\providecommand{\BIBentryALTinterwordspacing}{\spaceskip=\fontdimen2\font plus
\BIBentryALTinterwordstretchfactor\fontdimen3\font minus
  \fontdimen4\font\relax}
\providecommand{\BIBforeignlanguage}[2]{{%
\expandafter\ifx\csname l@#1\endcsname\relax
\typeout{** WARNING: IEEEtran.bst: No hyphenation pattern has been}%
\typeout{** loaded for the language `#1'. Using the pattern for}%
\typeout{** the default language instead.}%
\else
\language=\csname l@#1\endcsname
\fi
#2}}
\providecommand{\BIBdecl}{\relax}
\BIBdecl

\bibitem{GE2}
H.~Do, N.~Lee, and A.~Lozano, ``Line-of-sight {MIMO} via intelligent reflecting
  surface,'' \emph{IEEE Trans. Wireless Commun.}, vol.~22, no.~6, pp.
  4215--4231, Jun. 2023.

\bibitem{Holo-105}
T.~Gong, P.~Gavriilidis, R.~Ji, C.~Huang, G.~C. Alexandropoulos, L.~Wei,
  Z.~Zhang, M.~Debbah, H.~V. Poor, and C.~Yuen, ``Holographic {MIMO}
  communications: Theoretical foundations, enabling technologies, and future
  directions,'' \emph{IEEE Commun. Surv. Tutor.}, vol.~26, no.~1, pp. 196--257,
  First-quater, 2024.

\bibitem{Holo-4}
{\"O}.~T. Demir, E.~Bj{\"o}rnson, and L.~Sanguinetti, ``Channel modeling and
  channel estimation for holographic massive {MIMO} with planar arrays,''
  \emph{IEEE Wireless Commun. Lett.}, vol.~11, no.~5, pp. 997--1001, May 2022.

\bibitem{Holo-2}
A.~Pizzo, L.~Sanguinetti, and T.~L. Marzetta, ``Fourier plane-wave series
  expansion for holographic {MIMO} communications,'' \emph{IEEE Trans. Wireless
  Commun.}, vol.~21, no.~9, pp. 6890--6905, Sep. 2022.

\bibitem{guo-ICC24}
X.~Guo, Y.~Chen, Y.~Wang, Z.~Wang, and Z.~Han, ``Wavenumber domain sparse
  channel estimation in holographic {MIMO},'' in \emph{Proc. of IEEE ICC},
  Denver, CO, USA, Jun. 2024, available online:
  \url{https://arxiv.org/abs/2403.11071}.

\bibitem{Holo-2-OJCOM}
A.~Pizzo, L.~Sanguinetti, and T.~L. Marzetta, ``Spatial characterization of
  electromagnetic random channels,'' \emph{IEEE Open J. Commun. Soc.}, vol.~3,
  pp. 847--866, Apr. 2022.

\bibitem{Holo-15}
M.~Ghermezcheshmeh and N.~Zlatanov, ``Parametric channel estimation for {LoS}
  dominated holographic massive {MIMO} systems,'' \emph{IEEE Access}, vol.~11,
  pp. 44\,711--44\,724, May 2023.

\bibitem{chen-jsac3}
Y.~Chen, Y.~Wang, Z.~Wang, and Z.~Han, ``Angular-distance based channel
  estimation for holographic mimo,'' \emph{IEEE J. Sel. Areas Commun.},
  vol.~42, no.~6, pp. 1684--1702, Jun. 2024.

\bibitem{guo-tvt}
X.~Guo, Y.~Chen, and Y.~Wang, ``Compressed channel estimation for near-field
  {XL-MIMO} using triple parametric decomposition,'' \emph{IEEE Trans. Veh.
  Technol.}, vol.~72, no.~11, pp. 15\,040--15\,045, Nov. 2023.

\bibitem{sunshu}
Y.~Chen, R.~Li, C.~Han, S.~Sun, and M.~Tao, ``Hybrid spherical- and planar-wave
  channel modeling and estimation for terahertz integrated {UM-MIMO and IRS}
  systems,'' \emph{IEEE Trans. Wireless Commun.}, vol.~22, no.~12, pp.
  9746--9761, Dec. 2023.

\bibitem{Holo-3}
A.~Pizzo, T.~L. Marzetta, and L.~Sanguinetti, ``Spatially-stationary model for
  holographic {MIMO} small-scale fading,'' \emph{IEEE J. Sel. Areas Commun.},
  vol.~38, no.~9, pp. 1964--1979, Sep. 2020.

\bibitem{book-EM}
M.~Franceschetti, \emph{Wave theory of information}.\hskip 1em plus 0.5em minus
  0.4em\relax Cambridge University Press, 2017.

\bibitem{Holo_22-JSAC}
Y.~Liu, M.~Zhang, T.~Wang, A.~Zhang, and M.~Debbah, ``Densifying mimo: Channel
  modeling, physical constraints, and performance evaluation for holographic
  communications,'' \emph{IEEE J. Sel. Areas Commun.}, vol.~42, no.~6, pp.
  1504--1518, Jun. 2024.

\bibitem{Holo_27}
R.~Ji, S.~Chen, C.~Huang, J.~Yang, W.~E.~I. Sha, Z.~Zhang, C.~Yuen, and
  M.~Debbah, ``Extra {DoF} of near-field holographic {MIMO} communications
  leveraging evanescent waves,'' \emph{IEEE Wireless Commun. Lett.}, vol.~12,
  no.~4, pp. 580--584, Apr. 2023.

\bibitem{3GPP-38901}
``Technical specification group radio access network; study on channel model
  for frequencies from 0.5 to 100 {GHz (Release 17)},'' 3rd Generation
  Partnership Project, Tech. Rep. 3GPP TS 38.211 V17.5.0, Mar. 2022.

\end{thebibliography}

\section*{Biographies}

\textbf{Yuanbin Chen (M)} is currently a Post-Doctoral Research Fellow with the School of Electrical and Electronic Engineering, Nanyang Technological University (NTU), Singapore.
\\

\textbf{Xufeng Guo} is currently a Ph.D. candidate with the State Key Laboratory of Networking and Switching Technology, Beijing University of Posts and Telecommunications.
\\

\textbf{Gui Zhou (M)} is currently a Humboldt Post-Doctoral Research Fellow with the Institute for Digital Communications, Friedrich-Alexander-University Erlangen-N\"{u}rnberg (FAU), Germany.
\\

\textbf{Shi Jin (F)} is currently a Professor with the faculty of the National Mobile Communications Research Laboratory, Southeast University.
\\

\textbf{Derrick Wing Kwan Ng (F)} is now working as a Scientia Associate Professor at the University of New South Wales, Sydney, Australia. 
\\

\textbf{Zhaocheng Wang (F)} is currently a Professor with the Department of Electronic Engineering, Tsinghua University.

\end{document}